\documentclass{article}
\usepackage{psfig}
\title{Polchinski equation, reparameterization invariance and the derivative
  expansion}
\author{\\\\Jordi Comellas\thanks{Work supported by grants AEN95-0590
  (CICYT) and GRQ93-1047 (CIRIT).}\\\\
  \sl Departament d'Estructura i Constituents de la Mat\`eria\\
  \sl Facultat de F\'\i sica,\ Universitat de Barcelona\\
  \sl Diagonal, 647,\ 08028 Barcelona,\ Spain\\
  \tt comellas@sophia.ecm.ub.es\\\\\\}
\date{}
\begin{document}
  \begin{titlepage}
    \maketitle
    \thispagestyle{empty}
    \begin{abstract}
      The connection between the anomalous dimension and some invariance
      properties of the fixed point actions within exact RG is explored.
      As an application, Polchinski equation at next-to-leading
      order in the derivative expansion is studied.
      For the Wilson fixed point of the one-component scalar theory in three
      dimensions we obtain the critical exponents~$\eta=0.042$, $\nu=0.622$
      and~$\omega=0.754$.
    \end{abstract}
    \vfill
    UB-ECM-PF 97/05
  \end{titlepage}
  \section{Introduction}\label{sect.intr}
    The derivative expansion~\cite{rep_inv}
    is, up to now, the most reliable approximation
    available when dealing with exact renormalization group equations
    (exact RG equations, hereafter).
    Nevertheless, although quite simple at lowest order,\footnote{See,
    for instance, Ref.~\cite{lpa}.}
    some subtleties appear beyond it.

    They are mostly related to the preservation of the so-called
    reparameterization invariance~\cite{bell_wils,rep_inv_old}
    through the derivative expansion.
    In bold terms the problem is the following.  It is known
    that an overall
    normalization of the action\footnote{Hamiltonian,
    in condense matter language.}
    does not matter and, therefore, one may define RG transformations in
    such a way that its fixed points (FPs) are normalized to any
    pre-specified value.
    However, does this freedom survive order by
    order in the derivative expansion?  And if not, which is the ``best''
    normalization, in the sense that the expansion converges the most rapidly
    (if at all)?

    The problem is not merely academic because, as we will
    see, invariance of universal quantities under
    reparameterizations is the ultimate responsible of the uniqueness
    of the anomalous dimension $\eta$, much in the same way as the invariance
    under rescalings in a linear system over-determines its solution,
    effectively quantizing its eigenvalues.

    These two questions have been successfully answered for
    equations regarding the ``quantum effective action''~\cite{rep_inv}.
    Essentially, one chooses RG transformations in such a way that some
    invariance properties of the full equation are maintained order by order
    in the derivative expansion (see below for details).  In contrast,
    for the Polchinski equation~\cite{polchinski}, these points seem
    not even been addressed.

    Specifically, in Ref.~\cite{bhlm} a calculation of critical exponents
    for one-component scalar theories was presented, using Polchinski equation
    up to second order in the derivative expansion.  An arbitrary normalization
    of the action was chosen, and a large class of transformations analyzed.
    Critical indices
    were found to be, not surprisingly,
    transformation-dependent,
    as the arbitrary truncation of the expansion
    introduces spurious dependencies.
    These kind of ambiguities were partially
    solved by the usual criterion of minimal sensitivity~\cite{pms}, thus
    choosing the transformation which one hopes to converge most rapidly
    when derivative expanded.
    
    Two problems remain, nevertheless, open.  On one
    hand, as we have already mentioned, the possibility of non-standard
    normalizations was not addressed, nor a kind of reparameterization
    invariance sought.  On the other hand, a one-parameter family of
    transformations was left over
    for which, apparently, a minimal sensitivity criterion cannot be found.
    Different transformations within this line give slightly different
    RG eigenvalues although around a FP with significantly different
    anomalous dimensions.
    Thus one is left with the
    impression that clear results cannot be obtained
    and that the derivative expansion as it stands is thus
    of limited use beyond
    the leading order, at least for those type of equations.\hfil\break

    The first problem is specially annoying, because, although we expect
    that different transformations have different convergence properties and,
    thus, give different results, we do not expect that the {\em same\/}
    RG transformations may give a non-unique answer.

    The article is organized as follows.
    We first review in Section~\ref{sect.eq} the derivation of
    Polchinski equation together with its projection.
    The leading order of the derivative expansion is reviewed in
    Section~\ref{sect.lpa}.
    In Section~\ref{sect.eta} we explain the connection of the anomalous
    dimension~$\eta$ with the reparameterization symmetry and its relevance
    for next-to-leading order computations.
    A (failed) search for
    RG transformations invariant under a linear representation
    of the symmetry group is discussed in Section~\ref{sect.sym}.
    Finally, in Section~\ref{sect.res}, a detailed description of our method
    of calculation, together with the obtained results
    for critical indices~$\eta$, $\nu$ (given by the inverse of
    the relevant RG eigenvalue) and~$\omega$ (minus the first irrelevant
    RG eigenvalue) of the Ising
    universality class in three dimensions, is reported.
    Two peripheral subjects are left for the Appendices.
  \section{Equation}\label{sect.eq}
    Exact RG equations express the change of some
    quantity (usually the classical action) under an
    infinitesimal RG transformation.
    Polchinski equation deals with one-component bosonic models regulated as
    \begin{equation}
      S[\phi]=\frac{1}{2}\int_pp^2K^{-1}(p^2)\phi_{-p}\phi_p
      +S_I[\phi],
    \end{equation}
    with some function~$K(p^2)$
    decreasing faster than any polynomial for large~$p^2$.
    Conventions are as follows.  We work on Euclidean momentum space of
    dimension~$d$, with $\phi_p$~being
    the $p$~mode of the Fourier transformed field and
    \(
      \int_p\equiv\int\frac{d^dp}{(4\pi)^d}
    \).
    All variables are dimensionless.  Thus in dimensionful units the regulating
    function should be
    \(
      K(p^2/\Lambda^2)
    \),
    \(
      \Lambda\equiv\Lambda_0e^{-t}
    \)
    with $\Lambda_0$~some fixed scale and $t$~the label that parameterize
    the RG flow.

    RG transformations are usually defined in two steps, first some
    blocking---thinning of degrees of freedom---and afterwards a rescaling of
    variables so that FPs are possible.
    The first step is accomplished by splitting the field into
    \begin{equation}
      \phi_p=\phi^{(0)}_p+\phi^{(1)}_p,
    \end{equation}
    with the field~$\phi^{(0)}_p$ propagated by
    \begin{equation}
      \frac{K(p^2e^{2\tau})}{p^2}
    \end{equation}
    and $\phi^{(1)}_p$ by
    \begin{equation}
      \frac{K(p^2)-K(p^2e^{2\tau})}{p^2}=-2\tau K'(p^2)+{\cal O}(\tau^2).
    \end{equation}
    The field~%
    \(
      \phi^{(1)}_p
    \)
    is then to be integrated out, thus effectively eliminating modes
    mainly within the shell
    \(
      e^{-\tau}\!<|p|\le1
    \).
    For infinitesimal~$\tau$ the result is that functional integration
    over the full field~$\phi_p$ is equivalent to integration over~%
    \(
      \phi^{(0)}_p
    \)
    with the action
    \begin{equation}
      \frac{1}{2}\int_pp^2K^{-1}(p^2e^{2\tau})\phi^{(0)}_p\phi^{(0)}_{-p}
      +S_I
      +\tau\left\{
      \int_pS'_{I;-p}K'(p^2)S'_{I;p}
      -\int_pK'(p^2)S''_{I;-p,p}\right\},
    \end{equation}
    with
    \begin{equation}
      S_I\equiv S_I[\phi^{(0)}],\qquad
      S'_{I;p}\equiv\frac{\delta S_I}{\delta\phi_p}[\phi^{(0)}],\qquad
      S''_{I;-p,p}\equiv
      \frac{\delta S_I}{\delta\phi_{-p}\delta\phi_p}[\phi^{(0)}].
    \end{equation}
    See Ref.~\cite{integr} for further details.

    Rescaling of variables is carried out with
    \begin{equation}
      p\equiv\tilde pe^{-\tau},\ \ \
      \phi_p^{(0)}\equiv\phi_{\tilde p}e^{\tau\frac{d+2-\eta}{2}},
    \end{equation}
    where $\eta$ is some $t$-dependent parameter which coincides
    with the field anomalous dimension in the vicinity of a FP.

    Finally, renaming $\tilde p\rightarrow p$,
    \begin{eqnarray}\label{rge}
      \dot S[\phi]&\equiv&
        \lim_{\tau\longrightarrow0}\frac{S(t+\tau)-S(t)}{\tau}\nonumber\\
      &=&\int_pS'_{-p}K'(p^2)S'_p-\int_pK'(p^2)S''_{-p,p}
        -2\int_pp^2K^{-1}(p^2)K'(p^2)\phi_pS'_{I;p}\nonumber\\
      &&\mbox{}+dS-\frac{d-2+\eta}{2}\int_p\phi_pS'_p
        -\int_p\phi_p\,p\!\cdot\!\frac{\partial'}{\partial p}\,S'_p,
    \end{eqnarray}
    where we have (partially)
    returned back to the full action and written a prime in the
    last term to mean that it serves only to count momenta, thus
    delta functions of momentum conservation should not be
    derived.

    The derivative expansion amounts to consider a subset of interactions,
    with a maximum power of momenta in their coefficients, and to consider the
    RG Eq.~(\ref{rge}) for these terms as if they form a closed subset under
    the RG, that is, to approximate their evolution as if any other operators
    out of the subset do not contribute.

    We deal with the expansion up to two derivatives, with an action
    \begin{equation}
      S[\phi]=V[\phi_0]\int d^dx
      +Z[\phi_0]\,\frac{1}{2}\int_pp^2K^{-1}(p^2)\phi_{-p}\phi_p,
    \end{equation}
    which produces two coupled partial differential equations,
    \begin{eqnarray}\label{deriv_exp}
      \dot f&=&2K'(0)ff'-({\textstyle\int}K')f''-({\textstyle\int}p^2K')Z'
        +\frac{d+2-\eta}{2}\,f-\frac{d-2+\eta}{2}\,\varphi f',
          \nonumber\\\nonumber
      \dot Z&=&2K'(0)fZ'+4K'(0)f'Z+2K''(0)f'^2-({\textstyle\int}K')Z''
        -4K^{-1}(0)K'(0)f'\\
        &&\mbox{}-\eta Z-\frac{d-2+\eta}{2}\,\varphi Z',
    \end{eqnarray}
    with $\varphi \equiv\phi_0$ and $f(\varphi )\equiv V'(\varphi )$.
    For convenience, we work with such equations after the rescalings
    \begin{equation}
      \varphi \longrightarrow \sqrt{-{\textstyle\int}K'}\,\varphi ,\ \ \
      f\longrightarrow\frac{\sqrt{-{\textstyle\int}K'}}{-K'(0)}\,f,\ \ \
      Z\longrightarrow K^{-1}(0)Z;
    \end{equation}
    so that,
    \begin{eqnarray}\label{proj_eqs}
      \dot f&=&-2ff'+f''+AZ'
        +\frac{d+2-\eta}{2}\,f-\frac{d-2+\eta}{2}\,\varphi f',
          \\\nonumber
      \dot Z&=&-2fZ'-4f'Z+2Bf'^2+Z''
        +4f'-\eta Z-\frac{d-2+\eta}{2}\,\varphi Z',
    \end{eqnarray}
    where the transformation dependencies are bound to the parameters
    \begin{equation}
      A\equiv
        \frac{(-K'(0))(-{\textstyle\int}p^2K')}{(-{\textstyle\int}K')K(0)},
        \ \ \
      B\equiv\frac{K''(0)K(0)}{(-K'(0))^2}.
    \end{equation}
    These conventions coincide with those of Ref.~\cite{bhlm}, except
    for the arbitrariness of~$K(0)$.%
    \footnote{Concretely,
    after $Z\rightarrow1+2Z$ and $K(0)=1$ in Eq.~(\ref{proj_eqs}),
    one obtains Eq.~(5.1) of Ref.~\cite{bhlm}.
    The normalization $Z(0)=0$ chosen there corresponds
    thus to $Z(0)=1$ in the present
    paper.\label{comp.foot}}

    Eigenvalues of the RG control deviations from the FP.
    They can be calculated from the linearized version of the RG
    transformations.  That is, by taking
    $f(\varphi,t)=f_{FP}(\varphi )+re^{\lambda t}g(\varphi )$,
    $Z(\varphi,t)=Z_{FP}(\varphi )+re^{\lambda t}h(\varphi )$,
    with~$f_{FP}(\varphi )$ and~$Z_{FP}(\varphi )$ the FP values
    and $r\ll1$~a coupling constant,
    \begin{eqnarray}\label{eigen_eq}
      \lambda g&=&-2gf'_{FP}-2f_{FP}g'+g''+Ah'+\frac{d+2-\eta}{2}g
        -\frac{d-2+\eta}{2}\varphi g',
        \nonumber\\
      \lambda h&=&4Bf'_{FP}g'-4hf'_{FP}-4Z_{FP}g'-2h'f_{FP}-2Z'_{FP}g+h''
        +4g'\\\nonumber
      &&-\eta h-\frac{d-2+\eta}{2}\varphi h',
    \end{eqnarray}
    with $\lambda$ being the eigenvalue.
  \section{Local potential approximation}\label{sect.lpa}
    In this section we concisely discuss the results obtained within
    the first order of the derivative expansion---the
    so-called local potential approximation (LPA, hereafter).

    That is, we truncate the action simply to
    \begin{equation}\label{lpa_approx}
      S=\frac{1}{2}\int_pp^2K^{-1}(p^2)\phi_{-p}\phi_p+V(\phi_0)\int d^dx.
    \end{equation}
    The requirement that
    the non-potential term of Eq.~(\ref{lpa_approx})
    remains fixed under the RG forces $\eta=0$, while the evolution
    of $V(\varphi)$ is (cf.~Eq.~(\ref{proj_eqs}))
    \begin{equation}\label{lpa_eq}
      \dot f=-2ff'+f''+AZ'
        +\frac{d+2}{2}\,f-\frac{d-2}{2}\,\varphi f'.
    \end{equation}

    The FPs appear as the $\dot f=0$ solutions of Eq.~(\ref{lpa_eq}),
    \begin{equation}\label{lpa_fp}
      0=-2ff'+f''+AZ'
        +\frac{d+2}{2}\,f-\frac{d-2}{2}\,\varphi f'.
    \end{equation}
    Since this is a second order differential equation one may
    expect, in principle, a two-parameter set of FPs.
    This is not true, of course.  First of all, $Z_2$~symmetry
    imposes $f(0)=0$, thus reducing the arbitrariness to a uniparametric
    family of FPs (labeled by, for instance, $\gamma\equiv f'(0)$).
    Nevertheless, all but a finite number of~$\gamma$'s
    correspond to solutions
    which end with a singularity at a finite~$\varphi=\varphi_0$,
    \begin{equation}
      f(\varphi)\sim\frac{1}{\varphi_0-\varphi}\qquad
        \hbox{as $\varphi \to\varphi_0$}.
    \end{equation}
    Therefore, if we want~$f(\varphi)$ to be defined for the whole
    range~$0\le\varphi<\infty$,
    the possible FPs reduce to a finite set.

    The Gaussian FP corresponds to
    \begin{equation}
      f(\varphi )=0.
    \end{equation}
    It is easy to compute its eigenvalues because, after the rescaling
    $\varphi=2\tilde\varphi/\sqrt{d-2}$,
    the eigenvalue equation,
    \begin{equation}
      g''(\tilde\varphi)-2\tilde\varphi g'(\tilde\varphi)
        +2\frac{d+2-2\lambda}{d-2}g(\tilde\varphi)=0,
    \end{equation}
    is known to have polynomially bounded solutions, odd in $\tilde\varphi$
    (we want to maintain $Z_2$~symmetry), only for
    \begin{equation}
      \lambda=d-n(d-2),\qquad n=1,2,\ldots
    \end{equation}
    In this case the solutions are the well-known Hermite polynomials
    of odd degree~\cite{lpa}.
    The case $n=0$, $g(\varphi)=0$ corresponds to the
    identity operator (see Appendix~\ref{app__identity}).
    It is curious that this very same eigenvectors appear also
    within the LPA of other RG equations,
    although the aspect of the exact result
    may be quite different~\cite{hasen,lpa}.
    Being the FP action not a universal
    quantity, this behavior should be probably regarded just as a coincidence.

    One can also identify analytically a FP which must correspond to the
    non-critical high temperature FP,
    \begin{equation}
      f(\varphi)=\varphi.
    \end{equation}
    Its eigenvectors are also Hermite polynomials but without
    relevant directions,
    \begin{equation}
      \lambda=d-n(d+2),\qquad n=1,2,\ldots
    \end{equation}
    The case $n=0$ corresponds again to the identity operator.
    Because of the simplicity of the approximation, one can also
    explicitly write down the renormalized trajectory between the 
    Gaussian FP and this one here,
    \begin{equation}
      f(\varphi,t)=re^{2t}\varphi/(1+re^{2t}),
    \end{equation}
    with $r$ being some (fixed) coupling constant~\cite{lpa}.
    
    This latter FP is of no physical interest, but we
    wanted to quote it mainly because it has to be taken
    into account in the numerical work to follow, in order not to
    confuse it with the Wilson FP.

    Finally, there is also the FP corresponding to the Ising universality
    class, known as Wilson FP.  Unfortunately, one cannot obtain it
    analytically and must rely on numerical methods.

    The idea is to integrate Eq.~(\ref{lpa_fp})
    from initial conditions $f(0)=0$
    and~$f'(0)=\gamma$, and scan over different $\gamma$'s
    until the correct asymptotic behavior is reached,
    \begin{equation}
      f(\varphi)\sim \varphi+C\varphi^{\frac{d-2}{d+2}}+\cdots\quad
        \hbox{as $\varphi\to\infty$,}
    \end{equation}
    with $C$~being some arbitrary constant (which one must find out).
    Note that, as explained above, $C=0$ coincides with an uninteresting
    FP which should be discarded.
    For details about the numerical method we refer the interested
    reader to Appendix~\ref{app__num}.

    The eigenvectors are obtained in a similar manner.
    The linearized RG equation at this order is simply
    \begin{equation}\label{lpa_eigen}
      \lambda g=-2gf'_{FP}-2f_{FP}g'+g''+\frac{d+2}{2}g
        -\frac{d-2}{2}\varphi g'.
    \end{equation}
    In general its solutions will grow exponentially,
    \begin{equation}
      g(\varphi)\sim e^{\varphi^2}+\cdots\quad\hbox{as $\varphi\to\infty$,}
    \end{equation}
    whereas for a countable set of~$\lambda$'s it will be much smoother,
    \begin{equation}
      g(\varphi)\sim \varphi^{\frac{d-2-2\lambda}{d+2}}+\cdots
        \quad \hbox{as $\varphi\to\infty$.}
    \end{equation}
    Choosing the latter behavior, we obtain, from the first two eigenvalues,
    $\nu=0.649\,6$ and $\omega=0.655\,7$ in
    $d=3$.  See Refs.~\cite{bhlm,lpa}.

    Finally, we should mention that with our approach of focusing on~%
    $f(\varphi)\equiv V'(\varphi)$ instead of directly on~$V(\varphi)$
    we are always missing the identity operator.  A discussion about
    it is postponed to Appendix~\ref{app__identity}.
  \section{The anomalous dimension $\eta$}
    \label{sect.eta}
    In this section we explore
    the connection between the anomalous dimension~%
    $\eta$ and some exact symmetries
    of the FP action.

    There is a large class of RG transformations which are denoted as
    linear,\footnote{Not to be confused with
    linearized transformations around a given FP.} because
    the blocked variables are linearly
    related to the old ones \cite{bell1}.
    It includes the transformations expressed as exact
    differential equations.
    One of their main features is that
    they present always arbitrary parameters, usually in their rescaling part.

    These parameters are to
    be fine-tuned in order to obtain a FP.  For instance,
    the Gaussian FP of Polchinski equation,
    \begin{equation}
      \frac{1}{2}\int_pp^2K^{-1}(p^2)\phi_{-p}\phi_p,
    \end{equation}
    appears solely after the choice
    $\eta=0$.  If $\eta$ is set to some other value (in $d=4$), either
    no FP is obtained, or only a non-interacting, infinitely massive FP
    comes out, with no interest for Physics.

    A comment is in order.
    The operator~$\phi_0$ happens to be
    the first magnetic eigenvector, with
    eigenvalue $\lambda_M=(d+2-\eta)/2$.
    And, from the known scaling relation,%
    \footnote{See, for instance, Ref.~\cite{cardy}.}
    \begin{equation}
      \hbox{(anomalous dimension)}=d+2-2\lambda_M=\eta=0.
    \end{equation}
    Thus the free parameter~$\eta$ is directly the anomalous dimension.
    This
    is a general scenario, although the precise relation between the anomalous
    dimension and the free parameters is not always that simple~%
    \cite{wils_kog,bell_wils}.
    The key point, nonetheless, is that there always exists some parameters
    in the transformations themselves which need to
    be fine-tuned in order to reach a FP, and that these parameters are
    closely related to the field anomalous dimension at the FP.

    On the other hand, any FP is usually associated with
    a whole family of them, with different actions but with the same
    critical properties.  For instance, in the Gaussian case above,
    one finds the line of FPs,
    \begin{equation}\label{fp_line}
      \frac{1}{2}\int_pp^2K^{-1}(p^2)\left[1+aK(p^2)\right]^{-1}
        \phi_{-p}\phi_p,
    \end{equation}
    where $a$~is some (real) parameter with $aK(p^2)>-1$ for any~$p$.

    Of course, this is not a surprise, since it is precisely this arbitrariness
    what makes the set of FP solutions finite.
    It is like the scale invariance of a linear eigenvalue problem.
    The latter may have apparently a solution for each arbitrary eigenvalue,
    but the fact that we can choose the normalization of the eigenvector at
    will over-determines the system, making that only a discrete set of
    eigenvalues are allowed.
    The fine-tuning procedure, thus, is due to
    some sort of underlying arbitrariness
    (some sort of symmetry).
    Had a FP equation no invariances at all, no parameter
    would have to be fine-tuned.

    The form of the eigenperturbations about the FP also vary along the line
    of FP.  For instance, the mass term is
    \begin{equation}
      \frac{1}{2}\int_p[1+aK(p^2)]^{-2}\phi_{-p}\phi_p,
    \end{equation}
    with eigenvalue 2.

    Therefore, to summarize, linear RG transformations present two important
    features:
    They contain free parameters which have to be fine-tuned in order a FP
    to be reached; this feature reflects an underlying symmetry which
    manifest itself in a line of equivalent FPs.

    So far for exact computations.  Let us now discuss the actual
    case, that is, let us discuss how this picture survives under the
    (up to now) unavoidable approximations of RG equations.
    For definiteness, we take
    the derivative expansion at second order, Eq.~(\ref{proj_eqs}), but
    the following should be easily extendible to any feasible truncation.

    First of all,
    it is clear that the line of FPs, Eq.~(\ref{fp_line}),
    is hard to reproduce well:
    Each FP has different ${\cal O}(p^4)$~terms and higher and only
    those with such terms comparatively small will be correctly described
    by our truncation.  Therefore, one immediate consequence
    is that it is not that harmless to blindly choose
    one normalization (one value of $a$)
    of our FP action when some truncation
    is involved; e.g., in
    the Gaussian FP above, several choices of the parameter~%
    $a$ will be quite faithful while others will probably poorly capture
    the properties of the FP.

    Moreover, the opposite phenomenon also takes place: One
    also finds truncated FPs for the wrong choice of the transformation
    parameter~$\eta$.  To understand better this annoying feature,
    let us recall
    Wegner-Houghton equation~\cite{weg_hough,lpa},
    \begin{eqnarray}
      \dot S&=&\lim_{\tau\rightarrow0}\frac{1}{2\tau}
        \int_{e^{-\tau}<|p|\le1}\left[\ln S''_{-p,p}-
        S'_{-p}S'_p(S''_{-p,p})^{-1}\right]_{%
          \begin{array}{l}
            \scriptstyle\phi_p=0\\
            \raise 1ex\hbox{$\scriptscriptstyle e^{-\tau}<|p|\le1$}
          \end{array}
          }
        \nonumber\\
      &&\mbox{}+dS-\frac{d-2+\eta}{2}\int_p\phi_pS'_p-\int_p\phi_p\,
        p\!\cdot\!\frac{\partial'}{\partial p}\,S'_p.
    \end{eqnarray}
    A line of FPs is obtained here with
    \begin{equation}
      \int_{|p|\le1}\frac{d^dp}{(2\pi)^d}p^{2-\epsilon}\phi_{-p}\phi_p
    \end{equation}
    and $\eta=\epsilon$.
    This phenomenon is of quite different nature as the similar
    one described above.
    There the whole line shares the same critical properties,
    here it does not; there we have well-behaved actions throughout the
    fixed line, here nearly
    all of them are terribly non-local (in the sense that
    we cannot expand the action integrand in a power series of $p^2$).
    What happens is that we have one physical FP (the
    $\epsilon=0$ case)
    and a line
    of spurious ones.

    For exact calculations this is not a problem since RG transformations
    always map local actions into local actions and thus only
    the $\epsilon=0$ FP may be reached.
    Nevertheless,
    simply studying
    $V(\varphi )$ and~$Z(\varphi )$,
    one may not be able to distinguish non-local FPs
    from the local, physically meaningful, one.

    The recipe to cope these problems is not evident.  We have,
    nevertheless, some hints.  Clearly, one should succeed if one manages
    to compute a FP as local as possible (ideally, one FP with only the
    $V(\varphi )$ and~$Z(\varphi )$ terms).
    The problem is to try to reach this goal
    without computing the next order of ones expansion.
    The key is to try to check, within the given order,
    some known property of the exact solution,
    and take the approximate FP which best reproduces it.

    A glance at linear systems again may help.  Let us imagine that in an
    eigenvalue problem of linear algebra, one is so bold to approximate
    it in such a way that eigenvalues are no longer invariant under
    re-normalizations of the corresponding eigenvectors.
    One would probably find no quantization of eigenvalues.  However,
    if one insists, after all, to make sense of the approximation,
    a clever choice would be to look for the eigenvalues which are,
    to some extent, invariant under rescalings of their eigenvectors.
    This is, in fact, the method used from time to time for
    exact RG equations,
    following the procedure of Ref.~\cite{bell_wils,rep_inv_old}.

    The trick is to find the FP which is first-order (at least)
    independent of the free parameters.  This feature results
    in the presence of
    a redundant
    operator~\cite{wegner}
    which is marginal or nearly marginal.  It has to
    be redundant because it may not generate true RG trajectories, but
    simply reparameterizations of the theory; it has to be nearly
    marginal because it generates, at least approximately,
    a line of equivalent FPs (the equivalence being guaranteed by the
    redundancy).

    In our computations,
    therefore, we first plot the
    anomalous dimension versus the normalization of the kinetic term.
    It is seen that this curve presents always a maximum, which we regard as
    the best approximation for the actual anomalous dimension.
    This is further checked by the finding of a nearly marginal operator,
    which is known not to exist in the physical spectrum.  See Section~%
    \ref{sect.res} for the details.

    Before ending, let us make one further remark.
    The criterion explained so far
    relies on an invariance
    property of the physical, local, action.  This does not
    mean that non-local FPs do not present their own set of symmetries under
    reparameterizations.  But the claim is that the properties of local
    solutions should be well-described by local approximations.
    On the contrary, the analogous of the redundant marginal operator
    found for local FPs should be terribly non-local and, thus, poorly
    mapped within the truncation.  So poorly that they do not even show up~%
    \cite{bell_wils}.
  \section{Linearly symmetric RG transformations}\label{sect.sym}
    Before going on to actual computations, we would like to
    address one more issue.  One knows that
    the accurate anomalous dimension~$\eta$
    should be close to the one for which the action
    presents an (approximate) reparameterization symmetry.
    That is, why do we not choose RG transformations for which the symmetry
    is linearly realized, and try to keep this realization through
    the derivative expansion?
    This is, in fact, the approach taken in Ref.~\cite{rep_inv}
    for the effective-action type of exact RG equations.

    The search for such
    class of transformations
    is reported in this section.  The conclusions turn out
    to be, unfortunately, rather deceptive.  We find a class of
    regulators with a linearly
    realized reparameterization symmetry
    for the exact equations; but, first, the symmetry is broken at
    finite order in the derivative expansion; and, second, the regulators
    associated with such transformations turn out not to regulate, at least
    not in a finite order in the derivative expansion.\footnote{See
    \cite{dubna}, and references therein.}  Therefore, one is
    lead to the general scenario outlined in the previous section: One
    must scan a range of normalizations and decide on some kind
    of locality criteria which is the most reliable one.
 
    Let us seek some functions
    \(
      {\cal K}(\tilde p^2)
    \)
    and
    \(
      {\cal F}(\tilde p^2)
    \)
    such that the transformations
    \begin{equation}
      p=\lambda\tilde p,\qquad K(\lambda^2\tilde p^2)={\cal K}(\tilde p^2),
        \qquad
      \phi_{\lambda\tilde p}={\cal F}(\tilde p^2)\tilde\phi_{\tilde p}
    \end{equation}
    on an action converts it into another action that satisfies the same
    RG equation.%
    \footnote{One may try more general scenarios, like ${\cal F}={\cal F}
    (\tilde p,\tilde\phi_{\tilde p})$.  Nevertheless, not even in these general
    frameworks, any further invariant transformations have been found
    besides the ones considered in the text.}
 
    In order to obtain an action that satisfies again
    a Polchinski-kind equation,
    these functions must satisfy
    \begin{eqnarray}
      {\cal K}^{-1}{\cal K}'-{\cal F}^{-1}{\cal F}'&=&K^{-1}K'\nonumber\\
      \lambda^{-d-2}{\cal F}^{-2}{\cal K}'&=&K',
    \end{eqnarray}
    with solutions
    \begin{equation}
      {\cal K}(p^2)=\frac{\lambda^{\kappa}K(p^2)}{1+a(\lambda)K(p^2)},\qquad
      {\cal F}(p^2)=\frac{\lambda^{\frac{\kappa-d-2}{2}}}{1+a(\lambda)K(p^2)},
    \end{equation}
    for some function
    \(
      a(\lambda)
    \)
    and a real number $\kappa$.

    If we further require
    the RG transformations to be exactly the same as those of
    the initial action,
    the regulator must satisfy
    \begin{equation}
      K(\lambda^2p^2)=\frac{\lambda^{\kappa}K(p^2)}{1+a(\lambda)K(p^2)},
    \end{equation}
    which implies
    \begin{equation}\label{good_reg}
      K(p^2)=\frac{\kappa/\alpha}{1+\left(p^2/q^2\right)^{\kappa/2}},
    \end{equation}
    with
    \(
      \alpha\equiv-a'(1)
    \)
    and $q^2$, $\kappa$ real non-negative numbers.
    This are
    the regulators previously reported in Ref.~\cite{dubna}.
 
    If we were to use the above regulators, then, for every FP, there would
    exist a whole line of them, corresponding to different parameterizations
    of the action (different choices of the dummy variables $\phi$ and~$p$).
    The FPs within the line would be related by
    simple linear
    transformations, generated by a marginal redundant operator,
    as corresponds to any reparameterization.  All the FPs
    would be
    physically equivalent as long as infra-red properties are concerned~%
    \cite{wegner,bell_wils}.
    And if these RG transformations were used, then the discussion on
    normalizations would be totally void.
    Unfortunately, as we have pointed out
    above, there are important objections.
 
    First, this kind of symmetry
    does not in general hold order by order in the derivative expansion,
    unless~%
    \(
      a(\lambda)\equiv0
    \),
    as can be easily stated by considering Eq.~(\ref{deriv_exp}).
 
\begin{figure}[t]
      \centerline{\psfig{file=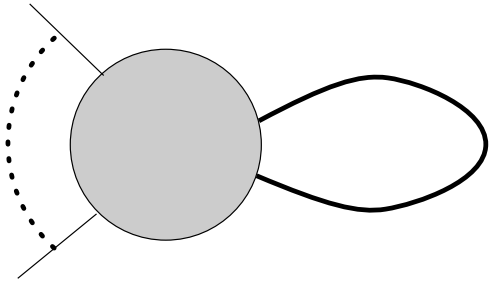}}
      \caption{Feynman
        diagram representing the second term in Polchinski equation.
        The shadow area stands for a generic vertex (which contributes
        with a coefficient of ${\cal O}(p^{2n})$ at $n$-th order in the
        derivative expansion and the thick solid line stands for a
        propagator~$\propto K'(p^2)$.  As usual one has to integrate over the
        closed loop.}
      \label{fig_onel}
    \end{figure}

    Second, the above class of functions~%
    \(
      K(p^2)
    \)
    do not regulate, at least they do not when we consider a fixed
    order in the derivative expansion, because they behave asymptotically
    as polynomials.
    That is, a simple study of one-loop diagrams (see Fig.~\ref{fig_onel})
    show, for instance,
    that the second term of the RG Eq.~(\ref{rge}) is not finite
    at~%
    \(
      {\cal O}(p^{2n})
    \)
    in the derivative expansion with~regulators of Eq.~(\ref{good_reg})
    whenever
    \(
      n\ge\kappa+1-d/2
    \).
 
    The conclusion is, therefore, that the symmetry cannot be
    linearly implemented order by order
    in the derivative expansion, in contrast to other
    equations~\cite{rep_inv} for which the maintenance of reparameterization
    invariance selects a preferred scheme.
    Polchinski equation in this sense is unfortunate: One cannot avoid
    scanning over different normalizations and selecting a criterion
    to discriminate among them.
    The true FP corresponds only to one of the possible normalizations; the
    rest should probably describe non-local actions.

    Finally, let us look again at the Gaussian FP, Eq.~(\ref{fp_line}).
    It can be found with a regulator of the form of Eq.~(\ref{good_reg}),
    and in such a case we have
    a nice example of the symmetry we have talked about.
    Our second objection does not apply here, since the Gaussian
    FP is ultra-violet
    finite anyhow.
    Nonetheless, the example shows that linear transformations are not the
    whole story, since, even for arbitrary choices of the regulating
    function~$K(p^2)$, the line of FPs exist and it has to be
    generated by non-linear transformations.
    This last possibility is what we are seeking for the Wilson FP
    in $d=3$.
  \section{Results}\label{sect.res}
    \subsection{Gaussian FP}
      We have already solved the FP equation within the Gaussian FP exactly,
      and also for the LPA.  Let us briefly
      review how it appears in second order in the derivative expansion.

      It consists solely on a kinetic term with a constant potential,
      that is,
      \begin{equation}
        f(\varphi)=0,\qquad Z(\varphi)=\frac{1}{1+aK(0)}.
      \end{equation}
      And, from the second equation in (\ref{proj_eqs}), $\eta=0$.

      The eigenvectors have a quite simple expression in the case $a=0$.
      Namely, there appear
      the same Hermite polynomials for~$g(\varphi)$, with $h(\varphi)=0$,
      obtained when dealing with the
      LPA, with eigenvalues
      $\lambda=d-n(d-2)$.

      There are also other eigenvectors with (for $a=0$) $g(\varphi)=0$ and
      the function~$h(\varphi)$ a solution of
      \begin{equation}
        h''(\tilde\varphi)-2\tilde\varphi h'(\tilde\varphi)
          -2\frac{2\lambda}{d-2}h(\tilde\varphi)=0,
      \end{equation}
      with $\tilde\varphi\equiv\frac{\sqrt{d-2}}{2}\varphi$.
      Again they are Hermite polynomials, this time of even degree, and their
      eigenvalues
      \begin{equation}
        \lambda=-n(d-2),\qquad n=0,1,\ldots
      \end{equation}

      The general form for these last perturbations of the FP action
      is, for the $n$-th such operator,
      \begin{equation}
        \frac{1}{2}\int dx\,(\partial\phi)^2\phi^{2n}(x)+\cdots,
      \end{equation}
      with the dots standing for terms with less number of fields.
      All of them are irrelevant but the one with $n=0$, which is a marginal
      redundant operator, $\frac{1}{2}\int dx(\partial\phi)^2$,
      that generates a line of equivalent FPs.
      It clearly corresponds to a change of normalization of the original
      FP action.

      The general case $a\ne0$ is technically a bit more complicated.
      The eigenvalue equations are now
      \begin{eqnarray}
        \lambda g&=&g''+Ah'+\frac{d+2}{2}g-\frac{d-2}{2}\varphi g',\\\nonumber
        \lambda h&=&\frac{aK(0)}{1+aK(0)}g'+h''-\frac{d-2}{2}\varphi h'.
      \end{eqnarray}
      The marginal redundant operator is still
      $g(\varphi)=0$, $h(\varphi)=\mbox{constant}$, of course.
      And the mass operator also conserves its form,
      $g(\varphi)=\varphi$, $h(\varphi)=0$,
      but it is clear that the tower of eigenvectors of the form of Hermite
      polynomials will not survive and a more complicate computation
      has to be done.
      For instance, the eigenvector with eigenvalue $\lambda=4-d$ (the
      ``$\int d^dx\,\phi^4(x)$'' operator) is, for~$d\ne4$,
      \begin{eqnarray}
        g(\tilde\varphi)
          &=&8\tilde\varphi^3-\left(12+6A\frac{aK(0)}{1+aK(0)}\right)
          \tilde\varphi,\\\nonumber
        h(\tilde\varphi)&=&\sqrt{d-2}\,\frac{aK(0)}{1+aK(0)}
          \left[6\,\tilde\varphi^2-3\left(1+\frac{A}{4-d}\frac{aK(0)}{1+
            aK(0)}\right)\right].
      \end{eqnarray}
      Clearly, the FP with~$a=0$ plays a special role
      in the derivative expansion as far as convergence properties are
      concerned.

      We do not discuss it any longer here,
      as we want only to look at it as
      an illustration of the kind of problems that might appear for the,
      more interesting,
      Wilson FP.
    \subsection{High temperature FP}
      The analysis of the high temperature FP is also easily extendible
      to one further order in the derivative expansion.  It appears as
      $f(\varphi)=\varphi$, as before, and $Z(\varphi)=\bar Z$,
      with $\bar Z$~a constant related with~$\eta$ by the equation
      \begin{equation}
        0=\frac{(2-\eta)^2}{2}B+2(2-\eta)-2(2-\eta)\bar Z-\eta\bar Z.
      \end{equation}

      Note that now $\eta$ is totally arbitrary and there
      is no evidence of a line of equivalent FPs.\footnote{We mean FPs
      with the same, e.g., $\eta$.}
      Furthermore, for some choice of $\eta$ we have $\bar Z=0$,
      thus making
      self-evident its physical interpretation of an infinitely massive
      theory with zero correlation length.\footnote{See,
      for instance, Ref.~\cite{tfp}.}
      If~$\bar Z\ne0$, the same physical interpretation shows up when
      considering dimensionful variables,
      \begin{equation}
        S=\frac{\Lambda^2}{2}\int_p\phi_{-p}\phi_p+\frac{\bar Z}{2}\int_p
  p^2\phi_{-p}\phi_p+\cdots,
      \end{equation}
      and letting $\Lambda\to\infty$.
    \subsection{Ising universality class: FP}
      We now discuss the Wilson FP in $d=3$.  First we sketch the
      numerical method to find it, leaving the details
      for Appendix \ref{app__num}.
      Then we turn to the normalization suitable for
      dealing with the reminiscence of the line of FPs present in the exact
      computation.  Finally, we asked ourselves about
      the best choice of RG transformations.
      The answer is given in terms of rapidity of convergence of the derivative
      expansion, proposing a criterium which extends that of
      Ref.~\cite{bhlm}.

      The FP action is the solution of RG Eqs.~(\ref{proj_eqs})
      with $\dot f(\varphi,t)= \dot Z(\varphi,t)=0$,
      \begin{eqnarray}\label{fp_eqs}
        0&=&-2ff'+f''+AZ'
          +\frac{d+2-\eta}{2}\,f-\frac{d-2+\eta}{2}\,\varphi f',
            \nonumber\\\nonumber
        0&=&-2fZ'-4f'Z+2Bf'^2+Z''
          +4f'\\
          &&\mbox{}-\eta Z-\frac{d-2+\eta}{2}\,\varphi Z',
      \end{eqnarray}
      and the boundary conditions
      \begin{eqnarray}\label{asympt}
        &f(0)=Z'(0)=0,&\nonumber\\
        &f(\varphi)\sim\frac{2-\eta}{2}\varphi+
          C\varphi^{\frac{d-2+\eta}{d+2-\eta}}+\cdots,
          \qquad\hbox{as $\varphi\to\infty$,}&\\
          \nonumber
        &Z(\varphi)\sim D+\cdots,\qquad\hbox{as $\varphi\to\infty$,}&
      \end{eqnarray}
      where $C$ and $D$ are constants (which have to be determined).
      The first two conditions come from imposing $Z_2$-symmetry,
      while the last two come directly from the FP Eqs.~%
      (\ref{fp_eqs}), once we require the solutions to exist
      for the whole range $0\le\varphi<\infty$.

      That is,
      from the asymptotic conditions we have three free parameters ($C$,
      $D$, $\eta$).
      The use of $Z_2$~symmetry imposes two constraints.
      Only after fixing one more condition, e.g.,
      $Z(0)=1$, the system is completely determined.
      One obtains, thus, (apart from the two special FPs discussed above)
      one approximation for the Wilson FP for any normalization~$Z(0)$.
      This is to be compared with the case where one manages to maintain
      explicitly a reparameterization symmetry~\cite{rep_inv}.
      In the latter case
      the invariance allows to arbitrary
      fix the remaining parameter to whatever value we choose,
      just as in a linear
      system one is allowed to fix the scale of the solutions.
      On the contrary, our results {\em depend} on this last parameter~%
      $Z(0)$.

\begin{figure}[t]
        \centerline{\psfig{file=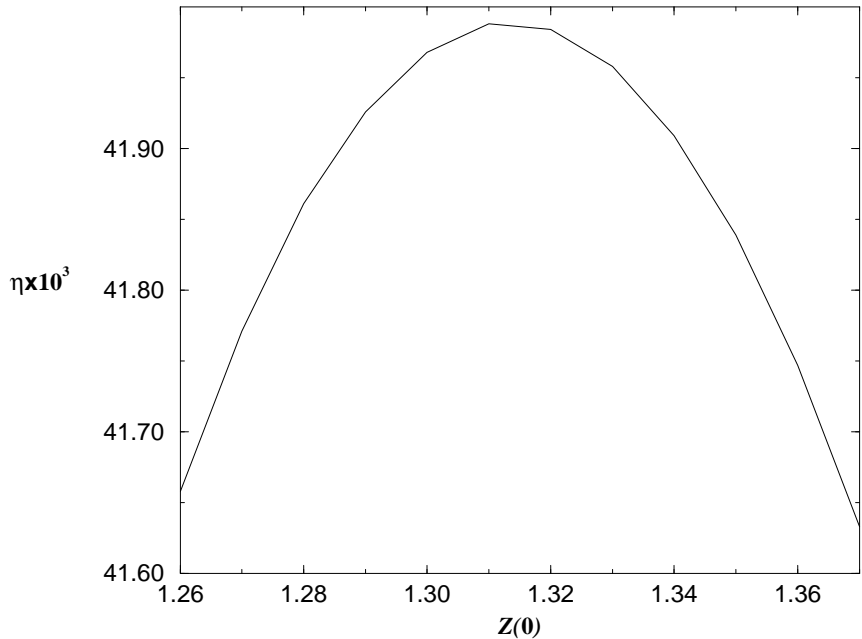,width=\textwidth}}
        \caption{Plot of the anomalous dimension $\eta$ for various FPs, with
          fixed $A=0.53$, $B=0.40$,
    as a function of their normalization $Z(0)$.
          These are
          the transformations we use later on to calculate eigenvalues.
          The maximum of the curve occurs at $Z(0)\sim1.31$ with $\eta=0.042$.}
        \label{fig__eta_v_z_2}
      \end{figure}

\begin{figure}[t]
        \centerline{\psfig{file=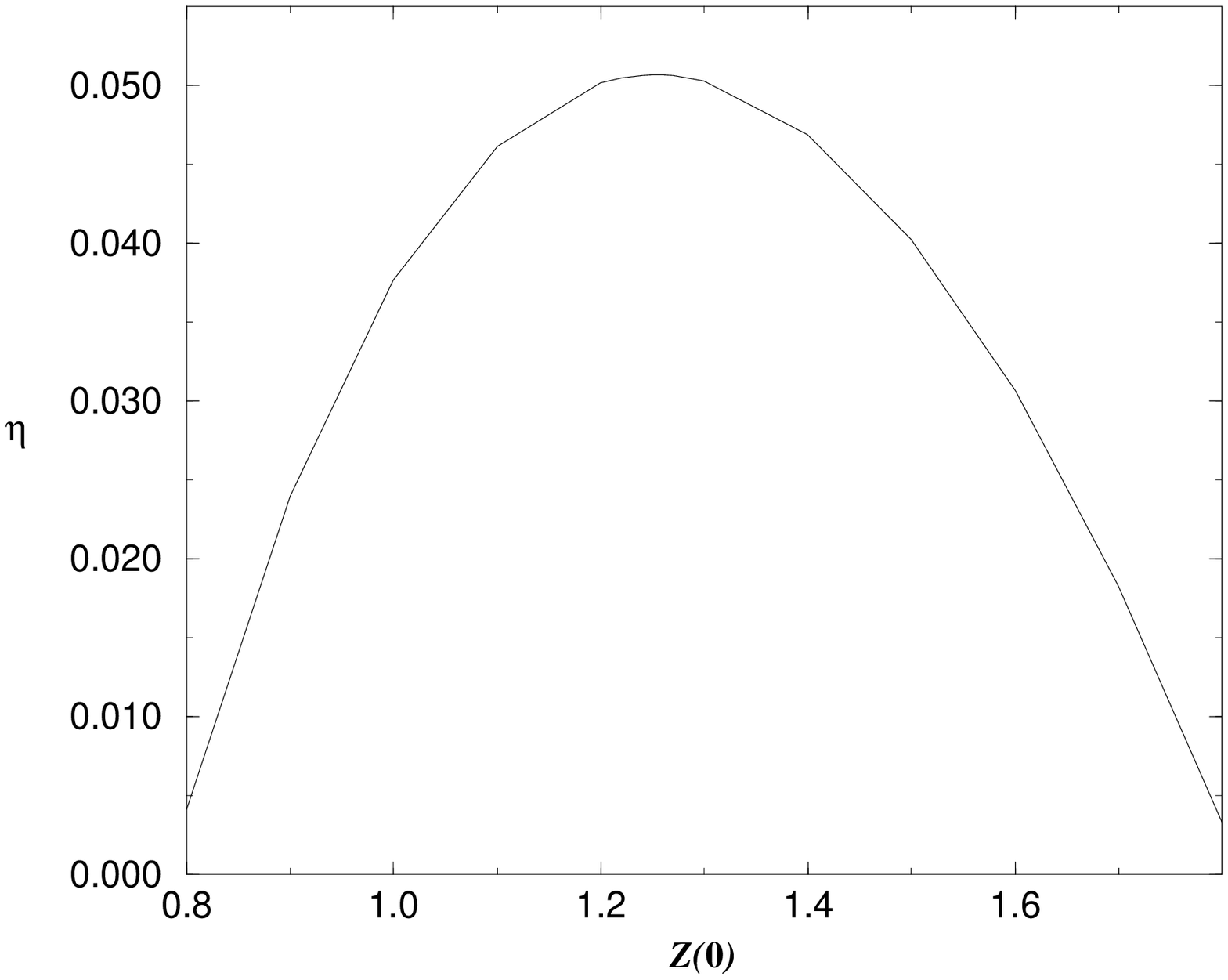,width=\textwidth}}
        \caption{Plot of the anomalous dimension $\eta$ for various FPs, with
          fixed $A=0.8$, $B=0.5$, as a function of their normalization $Z(0)$.
          These are the transformations used in Ref.~\protect\cite{bhlm}.
          The maximum of the curve occurs at $Z(0)=1.25$ with $\eta=0.051$.}
        \label{fig__eta_v_z_1}
      \end{figure}

      In Figs.~\ref{fig__eta_v_z_2},~\ref{fig__eta_v_z_1}
      we plot two examples.
      The first one is for $A=0.53$, $B=0.40$, which are the values
      that, as explain below, we take as the most reliable.  The second
      plot is for the values $A=0.8$, $B=0.5$ which are the ones taken in
      Ref.~\cite{bhlm}.  Note that the normalization chosen in that
      reference ($Z(0)=1$, in our notation) has nothing special from our
      point of view, while the appropriate normalization for which
      reparameterization invariance is (partially) recovered is $Z(0)=1.25$,
      with $\eta=0.51$.

      As explained earlier, this pattern is a consequence of the breaking
      of reparameterization symmetry.  The line of equivalent FPs is
      converted into a line of non-equivalent FPs because only the most
      local ones can be adequately described by our approximation, being
      the rest quite distorted.  Furthermore, the latter can be easily
      confused with some approximation of genuine non-local ones.
      And, as explained above, we take as a hint for locality the fact that
      a reminiscence of the line of FPs is present.  That is, we take as
      the closest to the exact Wilson FP the one with an anomalous dimension
      which is first-order independent of $Z(0)$.  For a given transformation
      (for fixed $A$ and~$B$) there is only one point with that property.

      Once we know how to compute the anomalous dimension given the
      RG transformations, we have to ask ourselves which are the
      most appropriate blockings.  This is an old subject, which has been
      studied mainly on the lattice.  For exact RG equations and
      their approximations it was addressed in the
      pioneering work of Ref.~\cite{bell_wils}.
      After it no much progress was achieved until~%
      Ref.~\cite{bhlm}, where a systematic search
      over different transformations
      was made and a proposal of ``best'' transformation was given,
      based on a reformulation of the principle of
      minimal sensitivity used in perturbation theory.

      Unfortunately, this latter reference is incomplete in two important
      directions, as we have already outlined in the introduction.
      The first one is its poor calculation of the anomalous dimension,
      without taking into account the problem of the breaking
      of reparameterization
      invariance of the theory, thus making their results not
      quite trustworthy.  The second problem is that their analysis did only
      half the way, leaving~$\eta$
      strongly dependent of one free parameter.

\begin{figure}[t]
        \centerline{\psfig{file=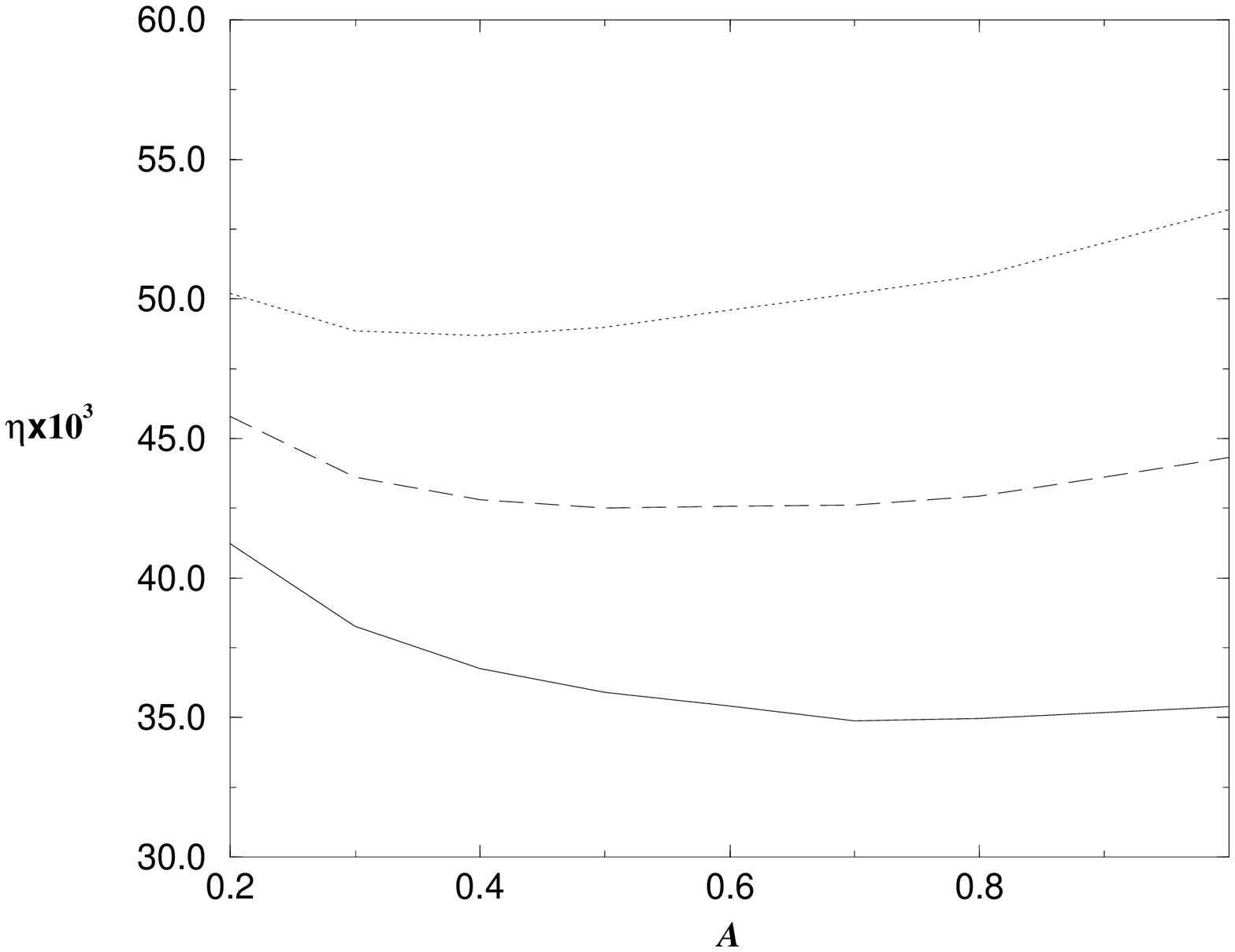,width=\textwidth}}
        \caption{The anomalous dimension~$\eta$ as a function of~$A$ for
          $B=0.30$ (solid line), $B=0.40$ (dashed line) and $B=0.50$ (dotted
          line).}
        \label{fig__eta_v_A}
      \end{figure}

\begin{figure}[t]
        \centerline{\psfig{file=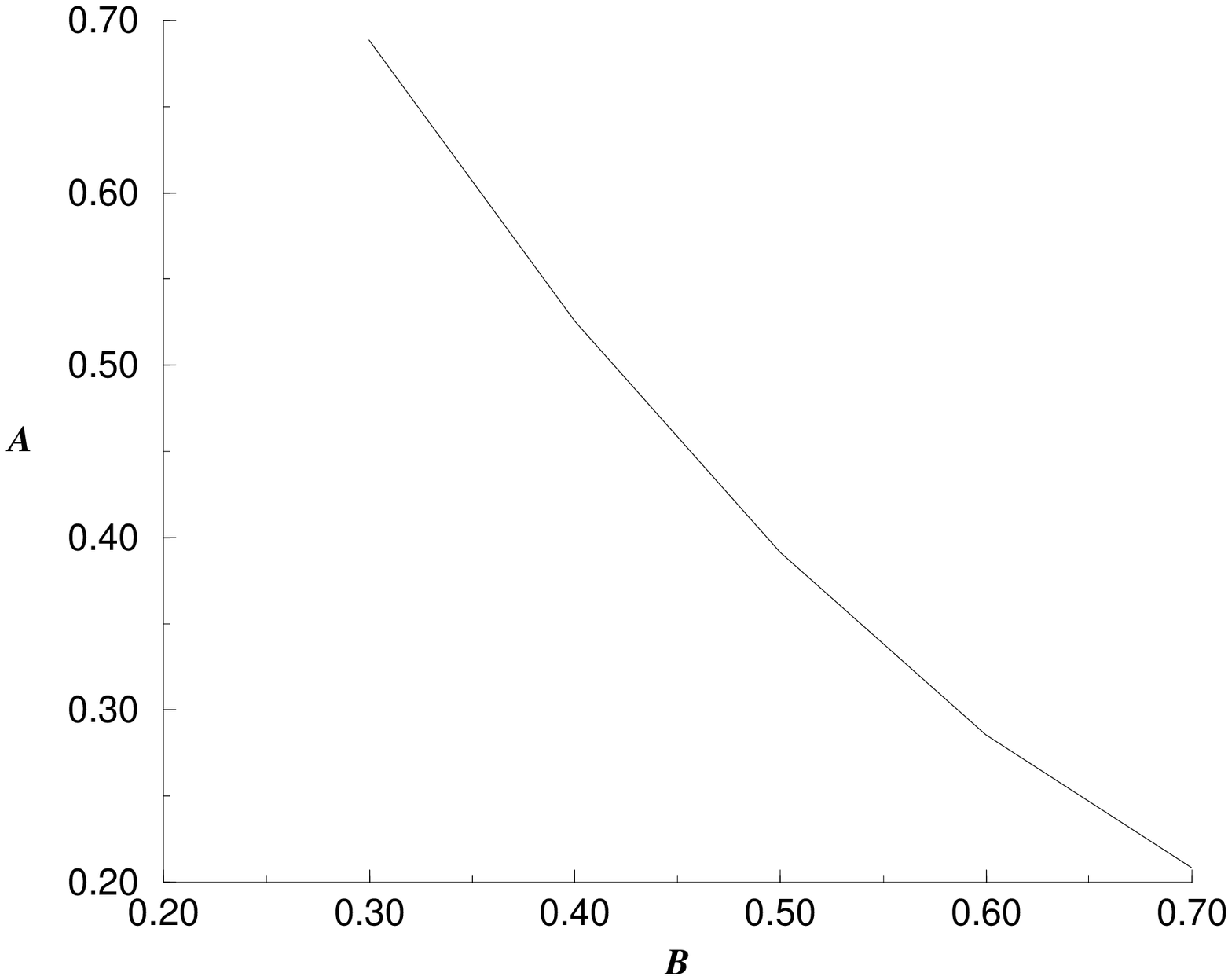,width=\textwidth}}
        \caption{The ``hollow'' of $\eta$ in the $A$--$B$ plane.}
        \label{fig__A_B}
      \end{figure}

      Nevertheless, we think the analysis of Ref.~\cite{bhlm}
      is basically correct,
      as far as the general dependencies on $A$ and~$B$ are concerned.
      That is, if a three dimensional plot of~$\eta$ as a function of $A$ and~%
      $B$ is made,
      one identifies a sort of ``hollow'' around which $\eta$ presents
      a one-dimensional minimum whenever the hollow is crossed.
      To plot it, it is best to fixed one parameter, say~$B$, and plot
      $\eta$ as a function of the other one, $A$~in our case.
      These kind of plots are shown in Fig.~\ref{fig__eta_v_A}, whereas a
      sketch of the location of the hollow in the $A$--$B$ plane is
      presented in Fig.~\ref{fig__A_B}.

      We choose, as in Ref.~\cite{bhlm}, a minimal sensitivity criterion
      to select the transformations for which $\eta$ yields
      on the hollow.  This should be the ones for which the derivative
      expansion converges the fastest.  See Ref.~\cite{pms} for an introduction
      of minimal sensitivity ideas.

\begin{figure}[t]
        \centerline{\psfig{file=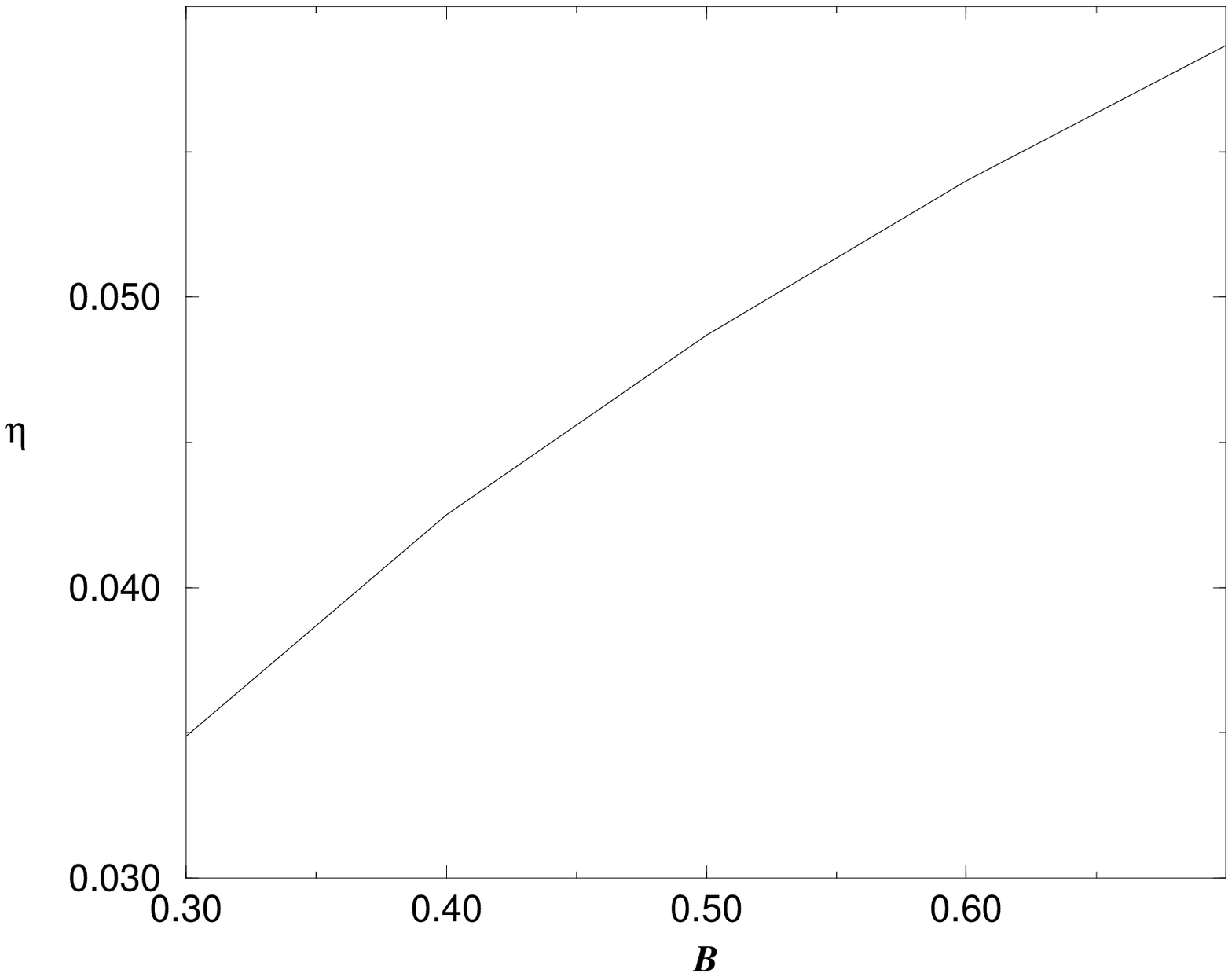,width=\textwidth}}
        \caption{The plot of $\eta$ along the hollow,
          parameterized by its $B$~value.
          This plot is to be compared with that of Fig.~\ref{fig__eta_v_A}.
          Here the curve is much steeper.}
        \label{fig__eta_v_B}
      \end{figure}

      Nevertheless, the strongest dependences of $\eta$ on the transformations
      occur {\em along} the hollow, not {\em across} it.  Compare, for
      instance, a plot of different $\eta$'s
      along the hollow (Fig.~\ref{fig__eta_v_B}) with Fig.~\ref{fig__eta_v_A}
      above.  In fact, the reason which prevents us to
      3d-plot $\eta$~versus the $A$--$B$ plane is that
      the quite different scales
      make it difficult to visually appreciate the details of the hollow.

\begin{figure}[t]
        \centerline{\psfig{file=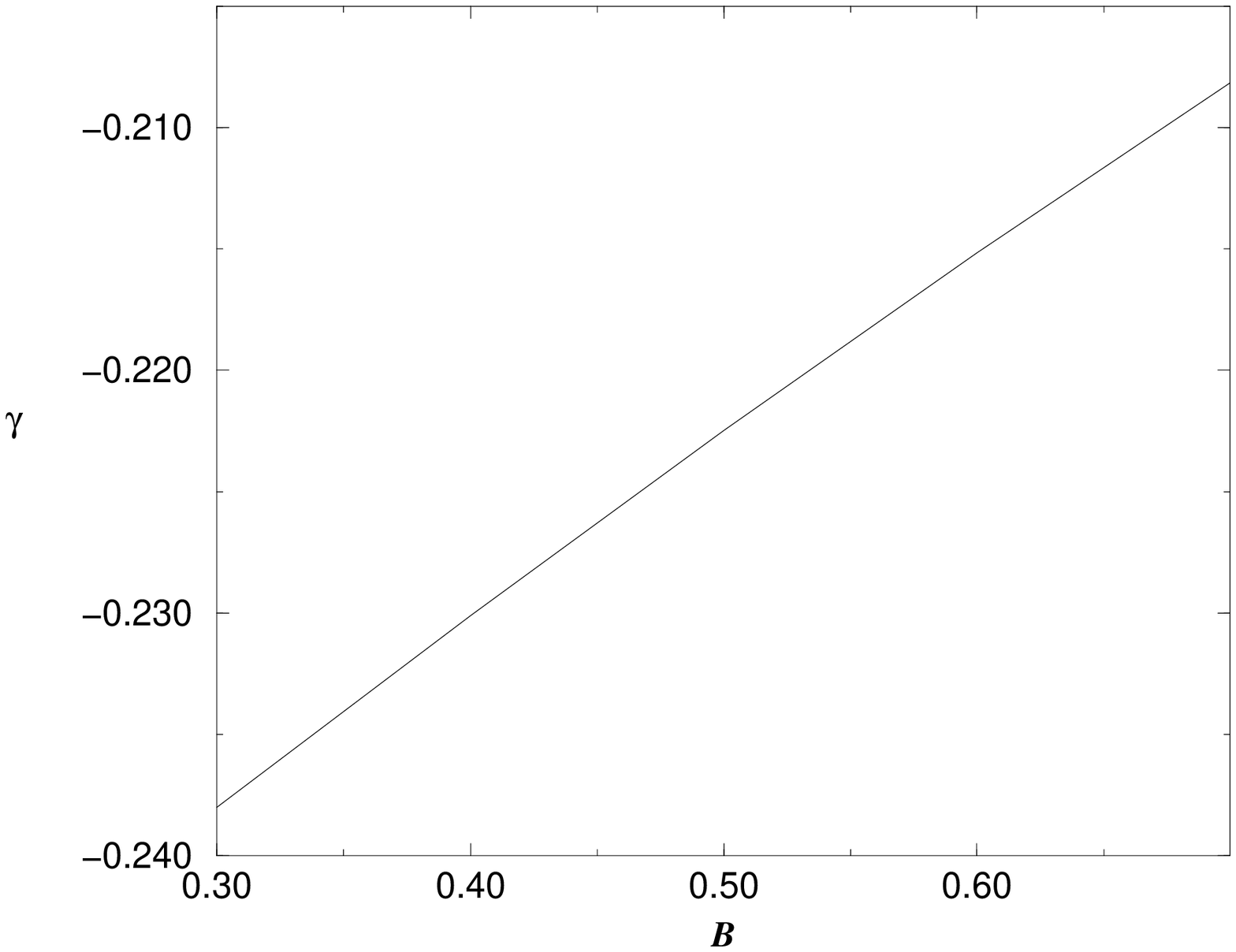,width=\textwidth}}
        \caption{The plot of $\gamma$ along the hollow,
          parameterized by its $B$~value.}
        \label{fig__gamma}
      \end{figure}

\begin{figure}[t]
        \centerline{\psfig{file=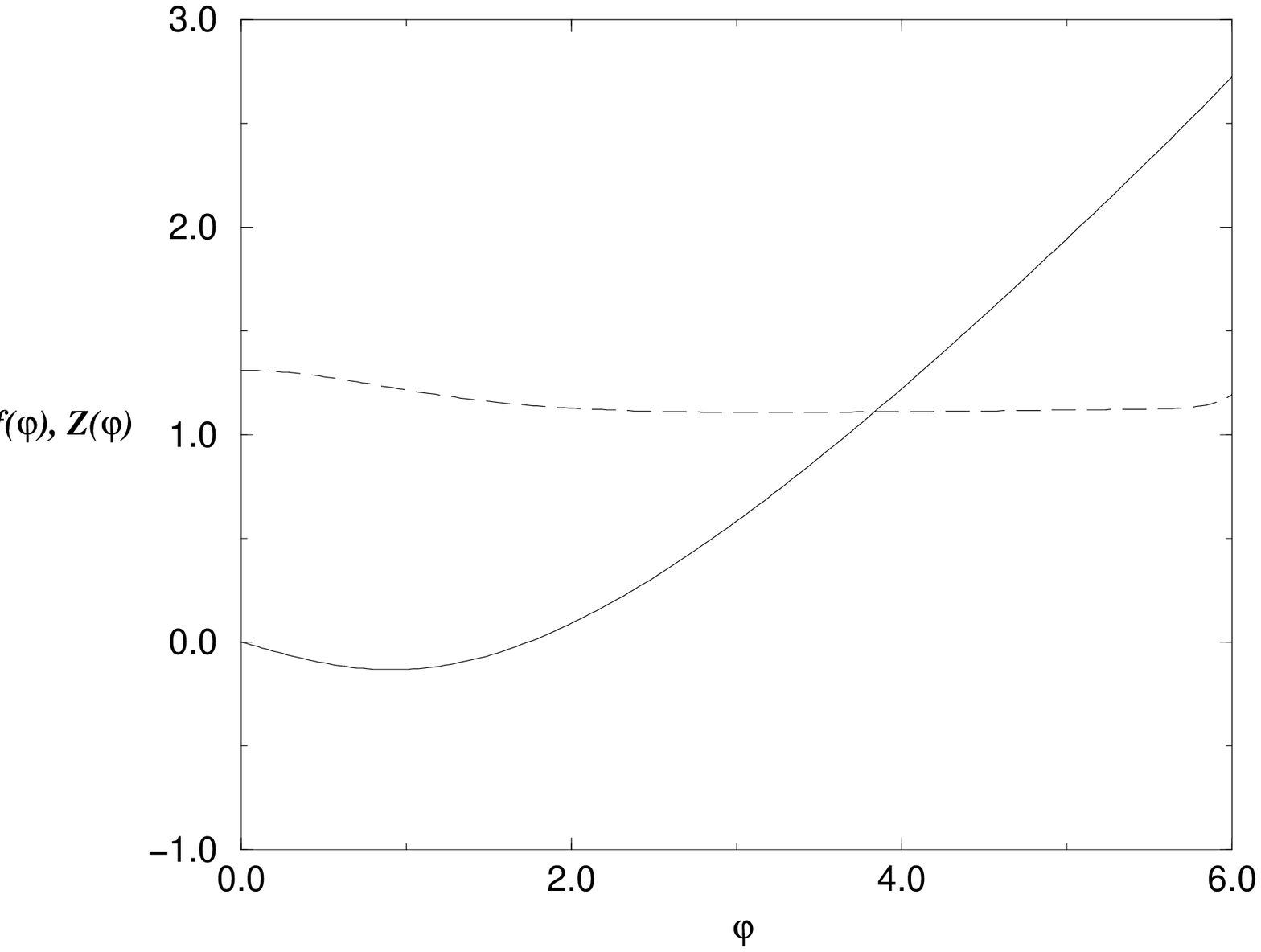,width=\textwidth}}
        \caption{The FP functions $f(\varphi)$ (solid line) and $Z(\varphi)$
          (dashed line)
    for second order in the derivative expansion with $A=0.53$,
          $B=0.40$ and $Z(0)=1.31$.}
        \label{fig__FP}
      \end{figure}

      We must then establish a criterion in order to keep only a small range
      of $B$~values if we wish to have a reasonable prediction for~$\eta$.
      If not, it can take nearly any value we want, since, as it was
      already observed
      in Ref.~\cite{bhlm}, it grows nearly linearly along the hollow.
      The other parameter one has to fine-tune to obtain the FP, e.g.,
      $\gamma\equiv f'(0)$, behaves also in a similar fashion
      (see Fig.~\ref{fig__gamma}).
      For consistency with previous choices,
      the criterion we choose must be based in convergence properties
      of the derivative expansion.
      Clearly, one cannot focus on $\eta$, since it changes from $\eta=0$,
      imposed by the LPA, to a value close
      (hopefully) to the actual one.
      This is because it has to do with the newly added term,
      $Z(\varphi)$, which was not present previously.  However, if
      the expansion converges sufficiently fast, the function $f(\varphi)$
      obtained at second order must not be too different from the one
      obtained at lowest order.
      This is the criterion we propose: One should
      trust those FPs whose values of
      $\gamma$ are close to the same, unique, number obtained
      from the LPA.
      This selects the transformations around those with $A=0.53$ and~%
      $B=0.40$, with $\gamma=-0.229\,0$ (to be compared with
      $\gamma=-0.228\,6$ from the LPA).
      Its anomalous dimension is $\eta=0.042$.
      The FP action is plotted in Fig.~\ref{fig__FP}.
      
\begin{figure}[t]
        \centerline{\psfig{file=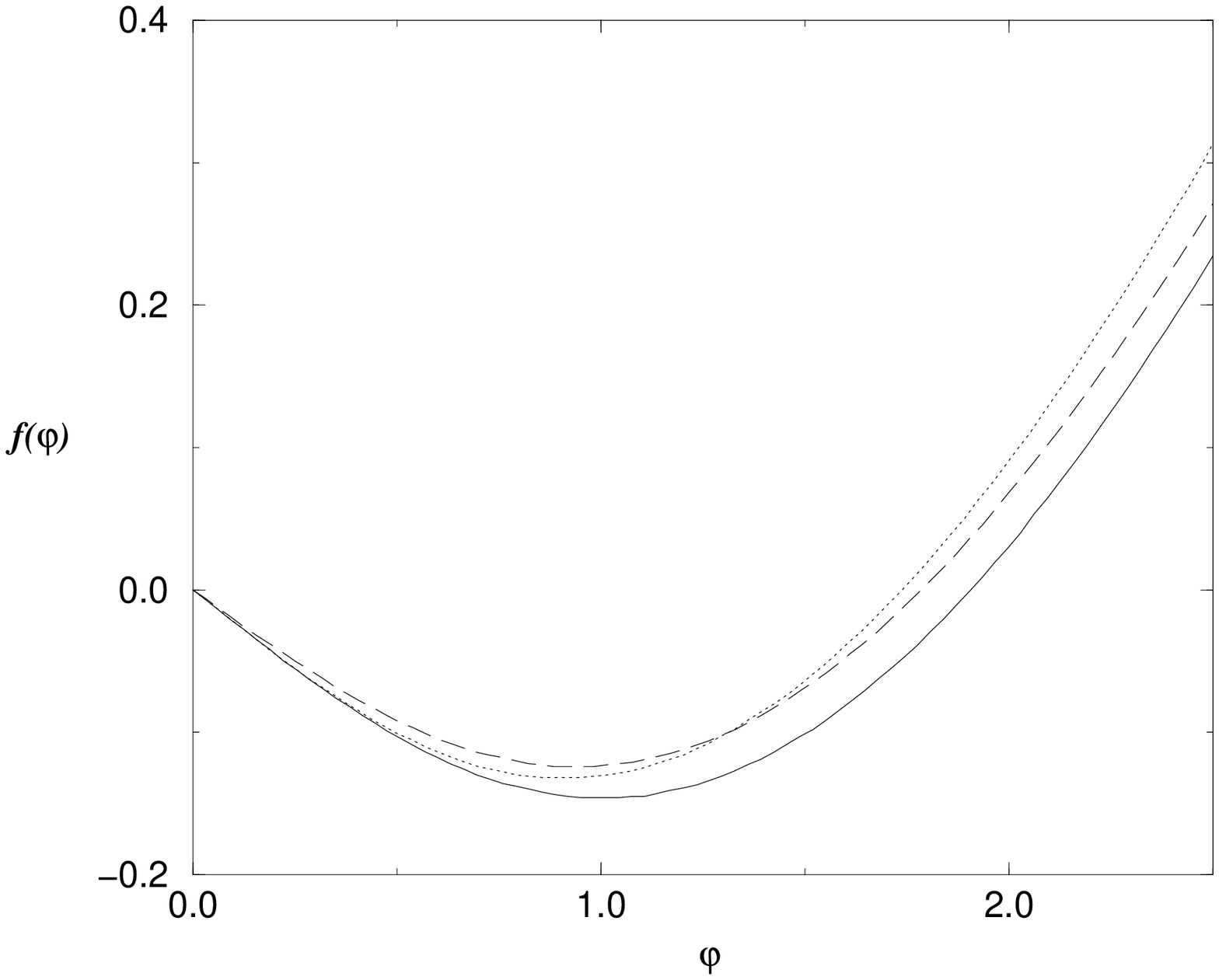,width=\textwidth}}
        \caption{Comparison of the function~$f(\varphi)$ from the
          LPA (solid line) with the second order
          ones corresponding to
          (1)~$A=0.21$, $B=0.70$ (dashed line), which becomes similar to the
          previous one for large~%
          $\varphi$, but differs significantly from it for~$\varphi\sim0$;
          (2)~$A=0.53$,
          $B=0.40$ (dotted line), which do not match with the first
          one for large~$\varphi$, but coincides with it quite well for~%
          $\varphi\sim0$.}
        \label{fig__comp}
      \end{figure}

      Note that with this choice we hope to reproduce well the
      behavior of the FP action near $\varphi=0$.  We do so following
      the common lore that this is the relevant part to reproduce
      if quantum fluctuations of the field are not very
      important, i.e., whenever $\eta$ is small.
      In fact, the whole idea would fail
      if the LPA was not sensible enough, which
      in turn is based on $\eta=0$ not being so crude an
      approximation.
      It is noteworthy, nonetheless, that there are other transformations
      which reproduce better than ours the asymptotic properties of
      $f(\varphi)$ but
      give quite worse results.  For instance, in Fig.~\ref{fig__comp}
      we plot the function $f(\varphi)$ for the LPA,
      together with the chosen $A=0.53$, $B=0.40$~FP and with that from
      $A=0.21$, $B=0.70$.  The last FP coincides quite well with the
      LPA
      one for~$\varphi\to\infty$ but
      differs from it significantly for~$\varphi\sim0$, just the
      opposite behavior of ours.  Nevertheless,
      our choice seem to give a better anomalous dimension ($\eta=0.042$
      instead of $\eta=0.058$).

      Before finishing this section, we would like to insist that we have
      just presented a standard calculation within the derivative expansion
      up to second order, without any further approximation.
      The discussion above is focused on the set of transformations best
      suited for such a calculation.  Without
      a clever choice, the expansion is poorly convergent and higher
      orders are necessary to provide the desired accuracy.
    \subsection{Ising universality class: critical indices}
      The computation of the eigenvalues is similar to that of the anomalous
      dimension $\eta$.  We have two asymptotic free parameters $g_1$ and~%
      $h_1$ for a
      polynomially growing $g(\varphi)$
      and~$h(\varphi)$,
      \begin{eqnarray}
        &g(\varphi)\sim g_1\varphi^{\frac{d-2+\eta-2\lambda}{d+2-\eta}}+\cdots
          \quad\hbox{as $\varphi\to\infty$,}&\\\nonumber
        &h(\varphi)\sim h_0\varphi^{\frac{-4+2\eta-2\lambda}{d+2-\eta}}
          +h_1\varphi^{\frac{-8+2\eta-2\lambda}{d+2-\eta}}+\cdots
          \quad\hbox{as $\varphi\to\infty$,}&
      \end{eqnarray}
      plus the eigenvalue $\lambda$ ($h_0$ is a parameter directly
      determined from $g_1$), but $Z_2$~symmetry
      imposes $g(0)=h'(0)=0$ and the linearity of the eigenvalue Eq.~%
      (\ref{eigen_eq}) allows us to fix on more parameter, say $g'(0)=1$.
      In this way the system gets determined and we encounter only a countable
      number of solutions.  We discuss the three most relevant ones.%
      \footnote{Recall
      that there is also the hidden solution~$\lambda=3$.  See Appendix~%
      \ref{app__identity}.}
      We concentrate here on the FP with $A=0.53$, $B=0.40$, $d=3$,
      as we stated above.

      The only relevant eigenvalue found gives $\nu=0.622$.

\begin{figure}[t]
        \vbox to 0sp{
          \centerline{\psfig{file=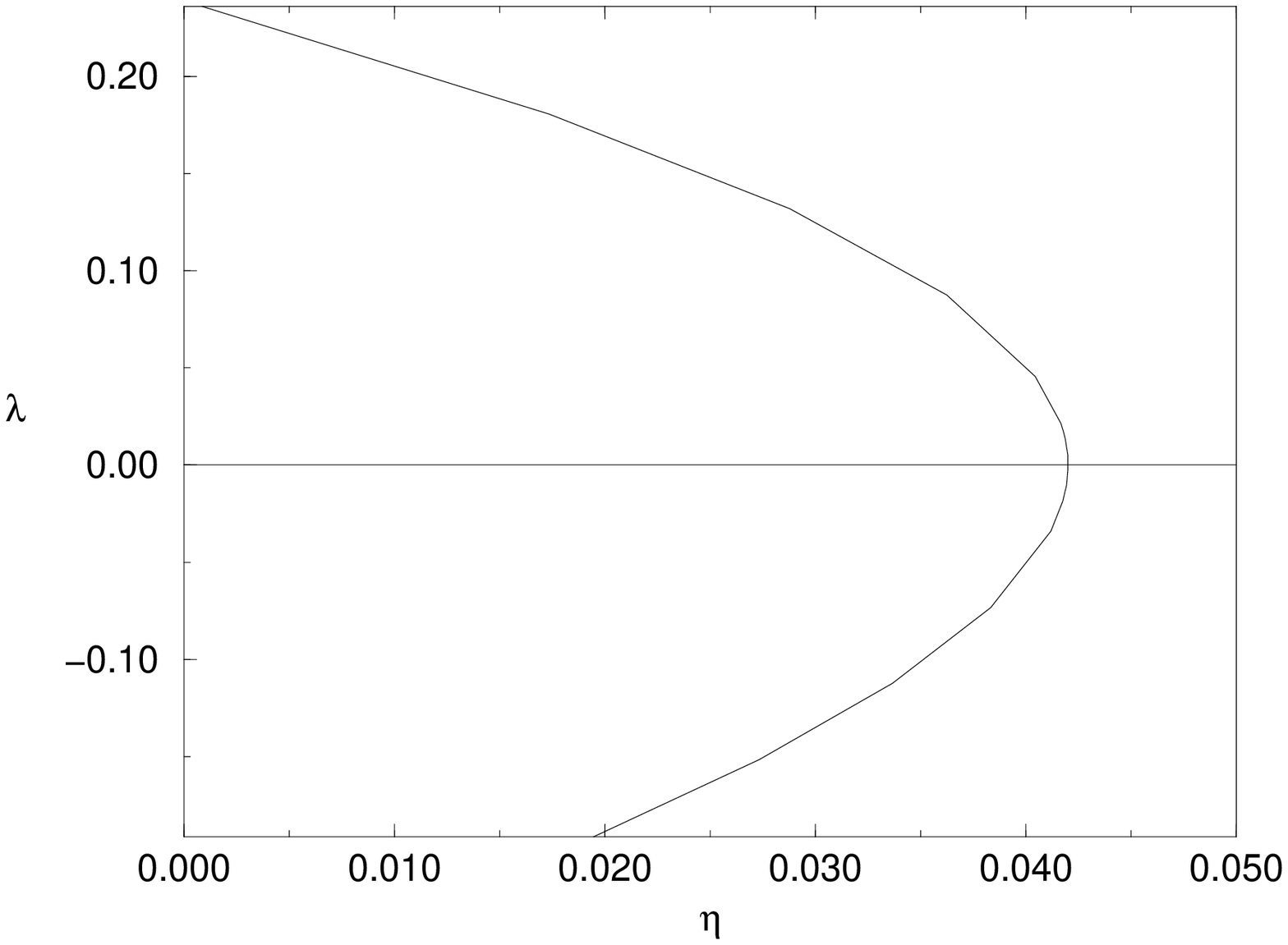,width=\textwidth}}
        \vss}
        \centerline{\psfig{file=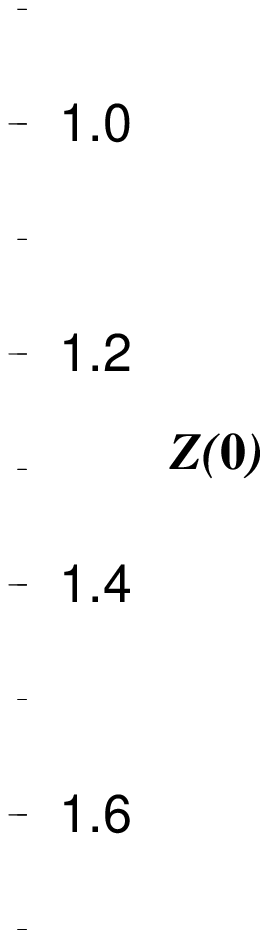,width=\textwidth}}
        \caption{The redundant eigenvalue~$\lambda$ as a function of~$\eta$
          for $A=0.53$, $B=0.40$ and varying~$Z(0)$.}
        \label{fig__marg}
      \end{figure}

      The second eigenvalue is not part of the physical spectrum, but
      corresponds to the redundant operator which generates the line of FPs
      in the exact case.
      In fact, its appearance is a crucial test of consistency of our
      method for calculating $\eta$.  This is so because
      it must necessarily
      be marginal, otherwise it cannot connect two points which both of them
      are FPs of the {\em same} RG transformations.
      For exactly $A=.53$, $B=0.40$ and $Z(0)=1.31$ we obtain $\lambda=0.001$.
      A plot of this redundant eigenvalue for different~$\eta$'s is given
      in Fig.~\ref{fig__marg}.  Note that for each~$\eta$ there are two
      values of~$\lambda$, corresponding to the two FPs which exist
      (cf.~Figs.~\ref{fig__eta_v_z_2}, \ref{fig__eta_v_z_1}).  The eigenvalue
      vanishes only when there is a unique FP given~$\eta$, i.e.,
      when $\eta$~is approximately invariant under~$Z(0)$.

      Finally, we seek for the first irrelevant eigenvalue, which controls
      the first deviations to scaling.\footnote{See again, for instance,
      \cite{cardy}.}  It gives $\omega=0.754$.

\begin{figure}[t]
        \centerline{\psfig{file=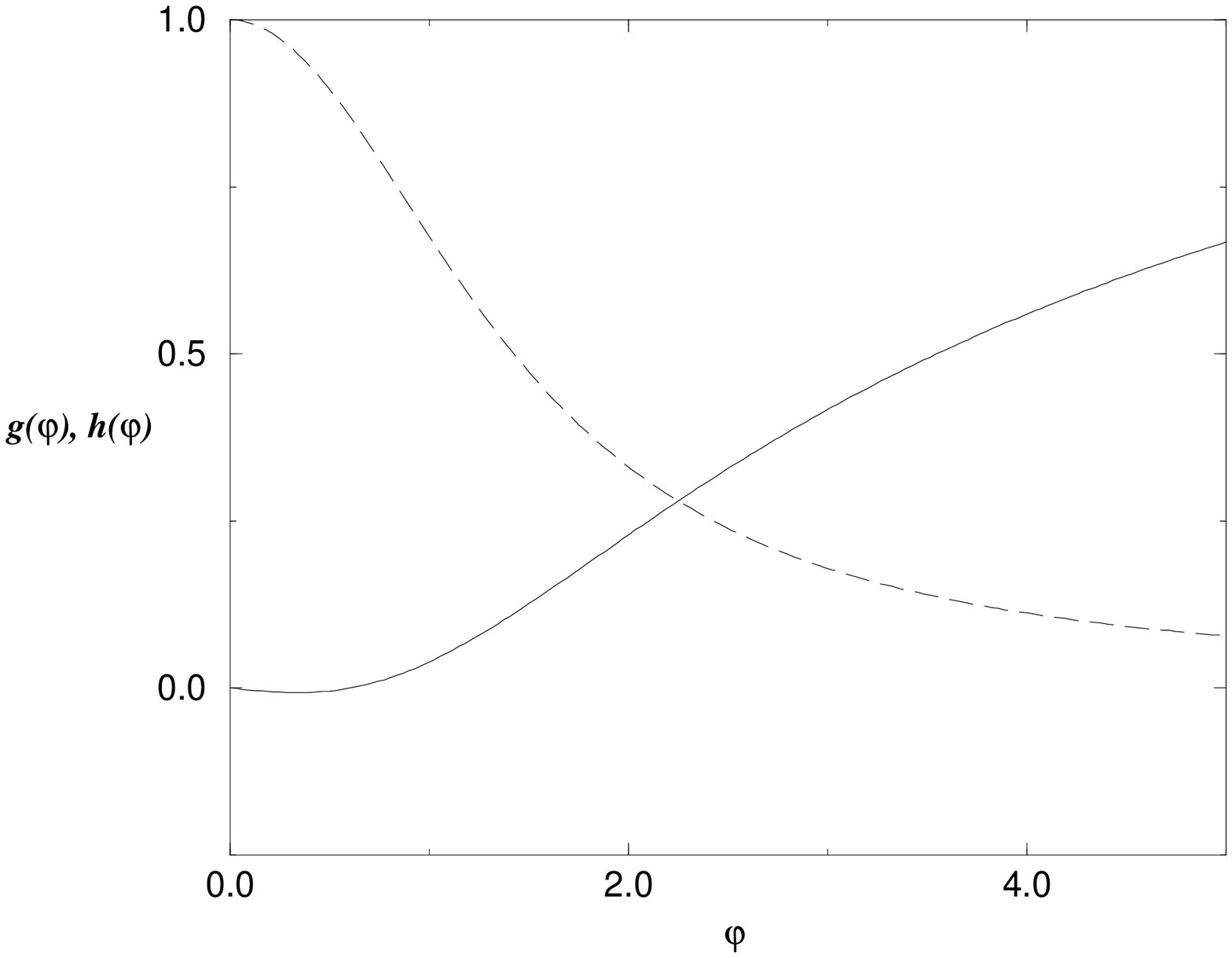,width=\textwidth}}
        \caption{The marginal redundant
          eigenoperator for $A=0.53$, $B=0.40$ and $Z(0)=1.31$.  The solid
          line corresponds to the function~$f(\varphi)$, while the
          dashed one to~$Z(\varphi)$.  For display purposes, we have chosen
          the normalization~$h(0)=1$, which gives $g'(0)=-0.031$.}
        \label{fig__eigen_mar}
      \end{figure}

      Together with the eigenvalues, there comes also their
      corresponding eigenvectors.
      As an example, we plot in Fig.~\ref{fig__eigen_mar}
      the one corresponding to the marginal redundant operator.
    \subsection{Accuracy}
      All figures given so far are significant.  Such a precision is achieved
      because, ones the problem has been set up, one can push the numerical
      calculations sufficiently far to obtain a high accuracy (of even five or
      six digits).  In fact, the main sources of errors are, on one hand,
      the problem introduced by
      different convergence properties of different transformations;
      and on the other, the poor knowledge of the corrections one may expect
      at higher orders.

      The first problem can be turn to our favor.  Once our criteria is
      accepted, there is only left the problem of to which accuracy we want~%
      $\gamma$ from the LPA be the same as the calculated at second order.
      Clearly, we expect the quantity to be corrected somehow
      at each order in the derivative expansion, so it is not sensible
      to ask for a terribly high coincidence.  If we restrict ourselves
      to FPs with~$\gamma$ three significant digits equal to the LPA
      value, then the anomalous dimension varies within the range
      $\eta=0.041\,9$--$0.042\,6$,
      thus providing us with an indication of the accuracy
      of our results.  The eigenvalues are, nonetheless, far more
      insensitive, specially the relevant one.
      As an extreme example take, for instance, $A=0.21$, $B=0.70$,
      which gives $\nu=0.631$, $\omega=0.689$.

      The main drawback of the whole framework, nevertheless, is its poor
      treatment of systematic errors introduced by the truncation of
      the expansion.  In fact, there are, up to now,
      no methods to estimate the contribution from higher orders.
      It is amazing that the same feature that makes the technique so powerful
      (its no use of a small parameter like~%
      $\epsilon\equiv4-d$ or the inverse number of fields~$1/N$),
      also prevents us to make a thorough study of its errors, thus reducing
      the reliability of its conclusions.

      Exact RG in this respect is special:\ in contradiction with what
      usually occurs, it happens to have the peculiarity that it is
      computationally quite simple, but its restrictions arrive in its
      deficient treatment of errors.  We have not a clear way to control the
      size of next corrections within the derivative expansion,
      nor wee can evaluate how good our criterium
      for improving convergence is.
      This is the main problem of the method, not the computation itself.%
      \footnote{Gauge theories are an exception. In this case the main
      problem is
      to find a non-perturbative gauge-invariant expansion to deal with
      the RG equation~\cite{gauge}.}
    \subsection{Discussion}
      Table~\ref{tab__sum} contains a summary of our results, together with
      a comparison with results from the LPA and also from other methods (exact
      RG for the effective action and a combination of best
      known estimates~\cite{rep_inv}).
      It is remarkable that with exact RG computations one may obtain
      numbers quite close to the best known
      estimates, and with considerably less
      effort.
      \begin{table}[t]
        \centerline{%
        \begin{tabular}{|c|c|c|c|c|}
          \hline
          &LPA&Polchinski&eff.\ action&best known\\\hline
          $\eta$&0&0.042&0.054&0.035(3)\\\hline
          $\nu$&$0.650$&0.622&0.618&0.631(2)\\\hline
          $\omega$&$0.656$&0.754&0.897&0.80(4)\\\hline
        \end{tabular}}
        \caption{The critical exponents~$\eta$, $\nu$ and~$\omega$ for
          (1)~the LPA of Polchinski equation; (2)~derivative expansion
          at second order of
          Polchinski equation; (3)~derivative expansion at second order of
          the effective action RG equation~\protect\cite{rep_inv};
          (4)~combination of best known estimates taken
          from Ref.~\protect\cite{rep_inv}.}
        \label{tab__sum}
      \end{table}

      We hope that the present paper serves to convince the reader that
      reported ambiguities in universal quantities computed with
      Polchinski equation~\cite{bhlm,fermions} do not constitute a problem,
      but rather can be turned to our favor.\footnote{See also
      Ref.~\cite{dubna}.}  There is still open, nonetheless, the problem
      of how one can compute systematic errors, which is the reason that
      prevents exact RG methods to be totally competitive with
      the best existing techniques (Monte Carlo RG~\cite{mcrg},
      $\epsilon$-expansion~\cite{eps}, \dots).

      As a final comment we want to recall that exact RG methods
      seem to give always an anomalous dimension~$\eta$ that increases
      from zero (LPA) to a value which is slightly above the best estimate;
      while the critical exponent~$\nu$ goes the other way round, going
      form a high value from the LPA to another one which is
      probably a bit low.  The exponent~$\omega$ does not seem
      to oscillate around its actual value, but this may only reflect our
      poor knowledge of it.  The apparent general behavior of $\eta$ and $\nu$
      is, however,
      quite intriguing, because if it is really general then it may constitute
      a first step towards an accurate estimation of errors.
  \section*{Acknowledgments}
    It is a pleasure to acknowledge interest and discussions with G.F. Golner,
    P. Hasenfratz, J.I. Latorre, A. Travesset and, specially,
    T.R. Morris, who has been saying much of this for some time.
  \appendix
  \section{Identity operator}\label{app__identity}
    It is known that, at any FP, there always exists an eigenoperator
    with eigenvalue $\lambda=d$,
    which is the highest dimensional operator
    of the theory.\footnote{Or the lowest dimensional
    operator in particle-physics
    language.  Recall that in condense-matter language one usually talks about
    the dimension~$D_{CM}$
    of an operator to refer to the (quantum) dimension
    of its coupling (in units of mass).  This is natural because these
    are the fields which appear in the partition function of the theory
    after the path-integration.  Particle physicists, however, usually
    talk about the dimension~$D_{PP}$ of the operator density itself, which is,
    trivially, $D_{PP}=d-D_{CM}$.  The identity operator thus have
    $D_{PP}=0$.}  Let us briefly review the arguments that support it~%
    \cite{ident}.

    Let us take any FP~action~$S^*$ and consider perturbations with a
    set of operators~${\cal O}_\alpha$,
    \begin{equation}
      S=S^*+j_\alpha{\cal O}_\alpha,
    \end{equation}
    where~$j_\alpha$ are some coupling constants.  Implicit summation over
    repeated labels is understood throughout.

    The partition function is then
    \begin{equation}
      Z=\int{\cal D}\phi\,e^{-S^*-j_\alpha{\cal O}_\alpha},
    \end{equation}
    and we define the thermodynamic densities
    \begin{equation}
      M_\alpha\equiv\frac{1}{V}\frac{\partial}{\partial j_\alpha}\ln Z,
    \end{equation}
    with~$V$ the volume of the system (needed in order $M_\alpha$~to be
    an intensive quantity and, thus, defined in the thermodynamic limit).

    After a RG~transformation, the system is shrunk in all linear
    dimensions and new couplings~$j'_\beta$ are introduced,
    \begin{equation}
      V'=e^{-d\tau}V,\qquad j'_\beta=j'_\beta(j_\alpha),
    \end{equation}
    with the partition function kept invariant.
    The above defined densities thus transform to
    \begin{equation}
      M_\alpha=e^{-d\tau}\frac{\partial j'_\beta}{\partial j_\alpha}
        \frac{1}{V'}\frac{\partial}{\partial j'_\beta}\ln Z
      \equiv e^{-d\tau}\frac{\partial j_\beta}{\partial j_\alpha}M'_\beta.
    \end{equation}

    Around a FP, the linearized RG~transformations are precisely the
    matrix
    \(
      {\cal R}_{\beta\alpha}\equiv\frac{\partial j'_\beta}{\partial j_\alpha}
    \),
    with the right hand side evaluated at the~FP.
    Therefore, using that, at a FP, $M'_\beta=M_\beta$,
    \begin{equation}
      {\cal R}_{\beta\alpha}M_\beta=e^{d\tau}M_\alpha.
    \end{equation}
    The final result is, therefore, that the densities~$\{M_\alpha\}$
    form an
    eigenvector of the linearized RG~transformations, with eigenvalue
    $\lambda=d$.\footnote{Again, the terminology is somewhat confusing.
    We have usually talked about eigenvectors to mean eigenoperators, that
    is, ${\cal O}=M_\alpha{\cal O}_\alpha$ in this case.}

    However, the above densities are expectation values of local
    operators and we may worry if all of them vanish.  That this
    is not the case is due to the existence of the
    identity operator, whose expectation
    value is
    \begin{equation}
      <\hbox{\bf 1}>\equiv\frac{1}{V}\frac{\partial}{\partial j_0}
        \ln\int {\cal D}\phi\,e^{-S^*+j_0V}=1.
    \end{equation}
    This completes the proof.

    With Polchinski equation, one may also trivially show that the identity
    operator is always an eigenvector with eigenvalue $\lambda=d$ (just
    substitute $S=S^*+\int d^dx$ in Eq.~(\ref{rge}) and use that $S^*$ is a
    FP).

    In the derivative expansion we do not notice it because it is $g(\varphi)=
    h(\varphi)=0$.  However, if the FP~potential is $V_{FP}(\varphi)$ and its
    eigenperturbations~$v(\varphi)$,
    then the linearized RG equation for the LPA may be rewritten as
    (cf.~Eq.~\ref{lpa_eigen})
    \begin{equation}
      \lambda v=-2v'V_{FP}'+v''+dv-\frac{d-2}{2}\varphi v',
    \end{equation}
    with~$\lambda=d$, $v=\hbox{constant}$, a trivial solution.  At next order
    we have to solve
    \begin{eqnarray}
      \lambda v&=&-2v'V_{FP}'+v''+Ah+dv-\frac{d-2+\eta}{2}\,\varphi v',
        \nonumber\\
      \lambda h&=&-2V_{FP}''h'-2v'Z_{FP}'-4V_{FP}''h-4v''Z_{FP}+4BV_{FP}''v''
        \\\nonumber
      &&\mbox{}+h''+4v''-\eta h-\frac{d-2+\eta}{2}\,\varphi h'.
    \end{eqnarray}
    And, again, a trivial solution is
    \begin{equation}
      \lambda=d,\qquad v=\hbox{constant},\qquad h=0.
    \end{equation}
  \section{Numerical methods}\label{app__num}
    We explain in this appendix the numerical methods used at second order.
    The calculation of the LPA results can be inferred form them and,
    moreover, they are discussed elsewhere~\cite{lpa}.

    The FP is searched by shooting from the origin to some point~$\varphi_0$
    where we impose the asymptotic conditions valid for~$\varphi_0\to\infty$,
    (cf.~Eq.~(\ref{asympt})),
    \begin{equation}
      \varphi_0\frac{f'(\varphi_0)-\frac{2-\eta}{2}}{
        f(\varphi_0)-\frac{2-\eta}{2}\,\varphi_0}
        =\frac{d-2+\eta}{d+2-\eta},\qquad
      Z'(\varphi_0)=0.
    \end{equation}
    Of course, these conditions cannot be imposed for finite $\varphi_0$, but
    we turn this fact to our advantage.  We move~$\varphi_0$ between 4 and~6,
    calculating $\eta$ and~$\gamma$ at each step and we observed that the
    results stabilizes for the five or six significant figures at some point
    between 5 and~6.  Thus it serves not only to check that the sub-leading
    asymptotic terms can be effectively discarded but also to check
    the accuracy of the overall numerical method.

    Incidentally, all numbers quoted are obtained always with at least one
    significant digit more than shown, to make sure that the numerical
    errors are under control.

    The computation is made with one normalization at a time, increasing it
    with finite amounts of~0.01.  From the FPs so obtained,
    we keep the one with maximum~$\eta$, as explained in
    Section~\ref{sect.res}.  We check that this grid does not
    affect the results, within our accuracy.  For instance, for the
    quoted~$\eta=0.042$, $A=0.53$,
    $B=0.40$, we have:\ for $Z(0)=0.30$, $\eta=0.041\,966$;
    for $Z(0)=0.31$, $\eta=0.041\,985$; and for $Z(0)=0.32$,
    $\eta=0.041\,981$.

    The scanning of the $A$--$B$ plane is made in a similar fashion (also
    with finite amounts of~0.01).  To identify the hollow mentioned in
    Section~\ref{sect.res} we fix first~$B$ and obtain the FPs for different~%
    $A$, and choose the one with the minimum $\eta$
    (see Fig.~\ref{fig__eta_v_A}).  Again, the grid should not be a great
    concern.  Compare, for $B=0.40$: $A=0.52$,
    $\eta=0.041\,985\,1$; $A=0.53$, $\eta=0.041\,984\,8$;
    $A=0.54$, $\eta=0.041\,989\,1$.

    Finally, the choice of a certain point within the hollow.
    This is a bit more difficult as computed quantities are far more sensitive
    to a variation within this line.  As explained in Section~\ref{sect.res},
    we take this
    as an indication of how far on can go with ones accuracy.
    We quote some representative numbers in Table~\ref{tab__acc}, which are
    to be compare with $f'(0)=-0.228\,60$ from the LPA.
    \begin{table}[t]
      \centerline{%
      \begin{tabular}{|c|c|c|c|c|}
        \hline
        $A$&$B$&$Z(0)$&$f'(0)$&$\eta$\\\hline
        0.55&0.39&1.31&$-0.230\,17$&$0.041\,3$\\\hline
        0.53&0.40&1.31&$-0.228\,99$&$0.042\,0$\\\hline
        0.52&0.41&1.32&$-0.228\,07$&$0.042\,6$\\\hline
        0.49&0.42&1.33&$-0.227\,11$&$0.043\,3$\\\hline
      \end{tabular}%
      }
    \caption{Comparison of different FPs on the ``hollow'' (see Section~%
      \ref{sect.res}).  Recall that $f'(0)=-0.288\,60$ within the LPA.}
    \label{tab__acc}
    \end{table}

    The computation of the eigenvalues is a quite easier problem, once the
    FP is determined (recall that now we are dealing with linear equations).
    Here we shoot from the origin (with $g'(0)$ fix) and fine-tune
    the initial condition~%
    $h(0)$ and the eigenvalue~$\lambda$, by imposing that the eigenfunctions
    $g(\varphi)$ and~$h(\varphi)$,
    together with their derivatives, remain bounded.
    (This somehow bold asymptotic conditions suffice.)
    Of course, one has to repeat the above analysis to be sure that
    the results are independent of the numerical method and, in particular,
    of the point where the asymptotic conditions are imposed.
  
\end{document}